# Exploring Large Language Models for Issuing Just-in-Time Adaptive Interventions:

## Fostering Physical Activity in a Conceptual Cardiac Rehabilitation Setting

David Haag[1,2,3], Devender Kumar[1,4], Sebastian Gruber[5,6], Dominik Hofer[6,7], Mahdi Sareban[1,8], Gunnar Treff[1,9], Josef Niebauer[1,8], Christopher Bull[10], Albrecht Schmidt[7], Jan David Smeddinck[1]

1. Ludwig Boltzmann Institute for Digital Health and Prevention, Salzburg, Austria

2. Department of Psychology, Paris-Lodron-University of Salzburg, Salzburg, Austria

3. Digital Health Information Systems, Center for Health & Bioresources, AIT Austrian Institute of Technology GmbH, Graz, Austria

4. University of Southern Denmark, Odense, Denmark

5. Institute of Business Informatics - Data & Knowledge Engineering, Johannes Kepler University Linz, Linz, Austria

6. Human Motion Analytics, Salzburg Research Forschungsgesellschaft, Salzburg, Austria

7. LMU Munich, Munich, Germany

8. University Institute for Sports Medicine, Prevention and Rehabilitation, Salzburg, Austria

9. Institute of Molecular Sports and Rehabilitation Medicine, Paracelsus Medical University, Salzburg, Austria

10. Open Lab, School of Computing, Newcastle University, Newcastle upon Tyne, UK

Corresponding author:

David Haag, Ludwig Boltzmann Institute for Digital Health and Prevention, Lindhofstr. 22, 5020 Salzburg, Austria. E-mail: david.haag@dhp.lbg.ac.at

**Abstract**

We evaluated the viability of using Large Language Models (LLMs) to trigger and personalize content in Just-in-Time Adaptive Interventions (JITAIs) in digital health. As an interaction pattern representative of context-aware computing, JITAIs are being explored for their potential to support sustainable behavior change, adapting interventions to an individual's current context and needs. Challenging traditional JITAI implementation models, which face severe scalability and flexibility limitations, we tested GPT-4 for suggesting JITAIs in the use case of heart-healthy activity in cardiac rehabilitation. Using three personas representing patients affected by CVD with varying severeness and five context sets per persona, we generated 450 JITAI decisions and messages. These were systematically evaluated against those created by 10 laypersons (LayPs) and 10 healthcare professionals (HCPs). GPT-4-generated JITAIs surpassed human-generated intervention suggestions, outperforming both LayPs and HCPs across all metrics (i.e., appropriateness, engagement, effectiveness, and professionalism). These results highlight the potential of LLMs to enhance JITAI implementations in personalized health interventions, demonstrating how generative AI could revolutionize context-aware computing.

# 1 Introduction

Supporting long-term health behavior change is challenging, especially when patients transition away from settings with intensive in-person support [14]. Sustaining adherence to healthier behavior becomes a key hurdle once direct human guidance is reduced, yet it is critical for long-term health outcomes. For instance, patients going through cardiac rehabilitation typically transition through different phases from initial clinical to ambulatory settings with fading healthcare professional (HCP) support and eventually should return to living fully independent everyday lives [44]. However, health behaviors such as regular physical activity (PA), which is a central part of a successful cardiac rehabilitation program [86], are strongly determined by contextual influences such as structural opportunities for integrating PA [37], self-regulatory capacity [26], or momentary affect [12, 20]. Thus, when HCP support fades after patients transition back to living their regular lives, many individuals quickly fall back into their previous, often sedentary habits [66]. At this stage, digital health elements such as Just-in-Time Adaptive Interventions (JITAIs) can play a crucial role in fostering and supporting the health behavior change introduced during the initial rehabilitation phases [61].

JITAIs are a concept that has many use cases across a broad range of digital health interventions [87]. Nahum-Shani et al. [51] conceptualized that JITAIs are designed to adjust to the dynamic needs and contexts of individuals, harnessing technological advancements to provide personalized health interventions when they are needed most. A simple example of a JITAI is to suggest a person to take the stairs instead of the elevator based on dynamic context information, such as location, when they enter a building. As opposed to common "one-size-fits-all" models in healthcare, JITAIs hold potential as a valuable tool offering more "precise", context-sensitive, and time-relevant health interventions that consider the individual's unique circumstances. However, the two main approaches currently used to implement JITAIs – *rule-based systems* and supervised *machine learning* (ML) models – are lacking behind visions of how flexible and effective JITAIs could be as a building block towards a "personal coach in your pocket" [3, 40].

JITAIs form an interaction pattern that implements principles of context-aware computing (CAC) [1], which has long been established as a relevant research area in HCI. This work presents an example of a broader "rethinking" of the original technical approaches to CAC. Traditionally, context-aware systems remained challenging to scale beyond clearly pre-defined contexts – or contextual information streams – informing closely coupled interactions. Complex and long-term personalization informed by multimodal contextual data streams was not attainable at scale with good resilience, reliability, acceptability, and user experience based on traditional approaches akin to those also employed with JITAIs.

Our main goal with this study was to explore the viability of employing LLMs, specifically ChatGPT based on GPT-4 as the top-performing model at the time of study execution, to trigger and generate JITAIs to support PA engagement in an outpatient cardiac rehabilitation setting. For that purpose, we presented the model with semi-structured information about a persona in outpatient cardiac rehabilitation together with a parameterization of their current context (see Supplement 1 for details). For any respective combination of persona and context, the model was tasked with (1) deciding whether the situation indicates a good opportunity for triggering a JITAI and, if so, (2) proposing a fitting motivational text for the potential use case of a push-message intervention. With this task, we aimed to investigate our guiding research question: "*Can LLMs be utilized as a decision-making and context-aware message tailoring mechanism in JITAIs for fostering PA support messaging in conceptual cardiac rehabilitation contexts?*"

To ground the performance of the LLM in relative terms, we compare the quality of GPT-4 powered ChatGPT (below abbreviated as GPT) generated JITAI decisions and message content to decisions and intervention messages generated by (1) *laypersons* (LayPs; as a "baseline" of human performance without domain-specific training and since it would at least appear procedurally possible to employ LayP for *crowdsourced* [33, 47, 62, 84] or *human computation* [48, 77] solutions for JITAI decision-making and generation in limited use cases), and (2) *HCPs* (as a "gold standard" – which we would not consider a practical alternative for solving the challenges of personalized JITAIs at scale, but can arguably provide valid anchoring of the LLM performance). To assess the decision and message quality of these three groups of "*generators*", we collected ratings on the dimensions of assumed appropriateness, engagement, effectiveness, and professionality of LayPs and HCPs acting as *"assessors"* (see Figure 2 for a visualization of the study flow).

Therefore, we explore the viability of contemporary generative AI (GenAI) systems, particularly LLMs, to integrate multimodal contextual data streams efficiently, to implement context-aware (just-in-time) adaptive interventions robustly and with good acceptability and user experience. We contribute (1) an approach for creating JITAI suggestions using LLMs and (2) empirical evidence that LLMs outperform human suggestions for personalized health interventions with regard to rater-assessed *appropriateness*, *engagement*, *effectiveness*, and *professionalism*. Our approach, outcomes, and findings can inform researchers, developers, and interaction designers in digital health and also form a use case of rethinking approaches from the established HCI concept of CAC based on the capabilities and potential of GenAI and particularly LLMs.

# 2 Background and Related Work

## 2.1 Traditional Context-Aware Computing

*CAC* is a fundamental aspect of pervasive and ubiquitous computing systems, where applications dynamically adapt to a user's situation by utilizing context information. Context is defined as any information that characterizes the situation of an entity, which can be a person, place, or object relevant to the interaction between a user and an application [1]. Since the early 1990s, context-awareness has enabled computing systems to provide task-relevant information and services, leveraging data from environmental sensors with minimal direct user input [59, 94]. This approach enhances human-computer interaction by adding meaningful context to computing processes, aligning with the pervasive computing vision of seamless integration into daily life [1, 68].

Development steps in traditional CAC focus on creating socio-technical systems that are not only aware of users' tasks but also understand their knowledge, backgrounds, and locations [16]. Ontological models can be used to organize and interpret sensor data into meaningful, higher-level concepts representing "situations", effectively allowing them to differentiate, e.g., between a user exercising or experiencing a medical emergency [75, 94]. This ability to differentiate between various situations and adjust communication and interactions accordingly also presents the fundamental requirement for effective JITAIs. However, at the current state, solely ontology-based systems are not able to deal with common challenges in open real-world settings. CAC systems have to manage vast quantities of (e.g., sensor) data while still ensuring accurate and precise adaptive behaviors [23], all whilst handling uncertainties around data completeness and quality. To this end, the integration of LLMs represents a promising avenue, as they possess the capacity to leverage prior knowledge and adapt to novel contexts through in-context learning and prompt engineering [36]. Nevertheless, the challenge persists in developing systems that are sufficiently flexible and robust to provide adaptive recommendations across a diverse range of goals, contexts, and user profiles [36].

## 2.2 Shortcomings of previous JITAI implementations

The limitations of traditional CAC also apply to previous JITAI implementations. Rule-based JITAIs [e.g., 10, 22], for example, function based on a pre-established set of rules or algorithms. They often implement clear-cut intervention guidelines that contain simple parameter thresholds. However, scalable and robust models handling the wider contextual dependencies of the targeted health behaviors are not yet available [12, 30]. Thus, setting up an effective set of rules requires time- and cost-intensive optimization processes, such as micro-randomized trials (MRTs) [38]. Additionally, these systems become exponentially more complex with rising numbers of tailoring variables. This makes them hard to manage and potentially leads to error-prone outputs [69]. Rule-based systems also do not efficiently scale to personalize interventions in the sense of tailoring to an individual's unique needs and preferences as they dynamically change over time [76].

Alternative approaches to such personalization, e.g., with ML models using reinforcement learning (RL) [24, 45], also face extensive requirements regarding the amount of data they need to start working effectively; cf. the "cold start" problem [78]. This can quickly leave users annoyed by ill-placed interventions, which may even cause abandonment of digital health tools [76]. Additionally, both rule-based and ML model-based JITAIs face limitations in effectively dealing with missing or sparse data or with not firmly pre-defined types of data [2], which are very common in real-world applications.

Furthermore, the concerns above are only considering JITAI tailoring in terms of whether a message should be sent at a given moment, not the content of messages, which is also a core intervention element with JITAIs. Currently, most state-of-the-art systems merely aim to support choosing the most appropriate option for a given situation out of a static list of intervention options with only heavily pre-structured or no contextualization and personalization. This limits adaptability and can quickly lead to JITAIs being repetitive, potentially disrupting the user's impression of receiving personalized support and possibly leading to reduced efficacy and increasing disengagement [29]. Overcoming these limitations arguably requires a different approach that offers the opportunity to generate flexible, personalized, context-aware, and ever-unique health interventions in real-time. Such qualities might be offered by GenAI, including large-language models (LLMs) [17, 63, 83]. This begs the question of whether these models might present a viable solution for the challenges named above and thus be the next step in the evolution of JITAIs.

## 2.3 LLM-based JITAIs

As evidenced by the high average quality of single-shot responses produced by leading foundation models [56], LLMs such as OpenAI's GPT series can generate meaningful responses when faced with opportunistically sampled, very high-dimensional information with many sparse or temporarily unavailable parameters. The adaptability, scalability, and potential for personalization [46] of LLMs could move JITAI systems (cf. Figure 1 for an illustration of the role of LLMs in terms of the original JITAI conceptualization [51]) closer to enabling the envisioned "personal coach in your pocket" concept. In the original JITAI concept, decision rules are evaluated at predefined decision points. This means that based on the values of one or more tailoring variables, a decision is made on whether one of the predefined intervention options should be triggered. These interventions aim to improve on a certain proximal outcome (e.g., step count within 30 minutes after intervention), which should, in turn, lead to improvements in a distal outcome (e.g., overall PA level). We propose that LLMs could (1) replace the decision rules in this classic JITAI conceptualization, offering more flexibility in the tailoring of interventions, and (2) augment the intervention options with custom-generated or more strongly adapted and context-aware content.

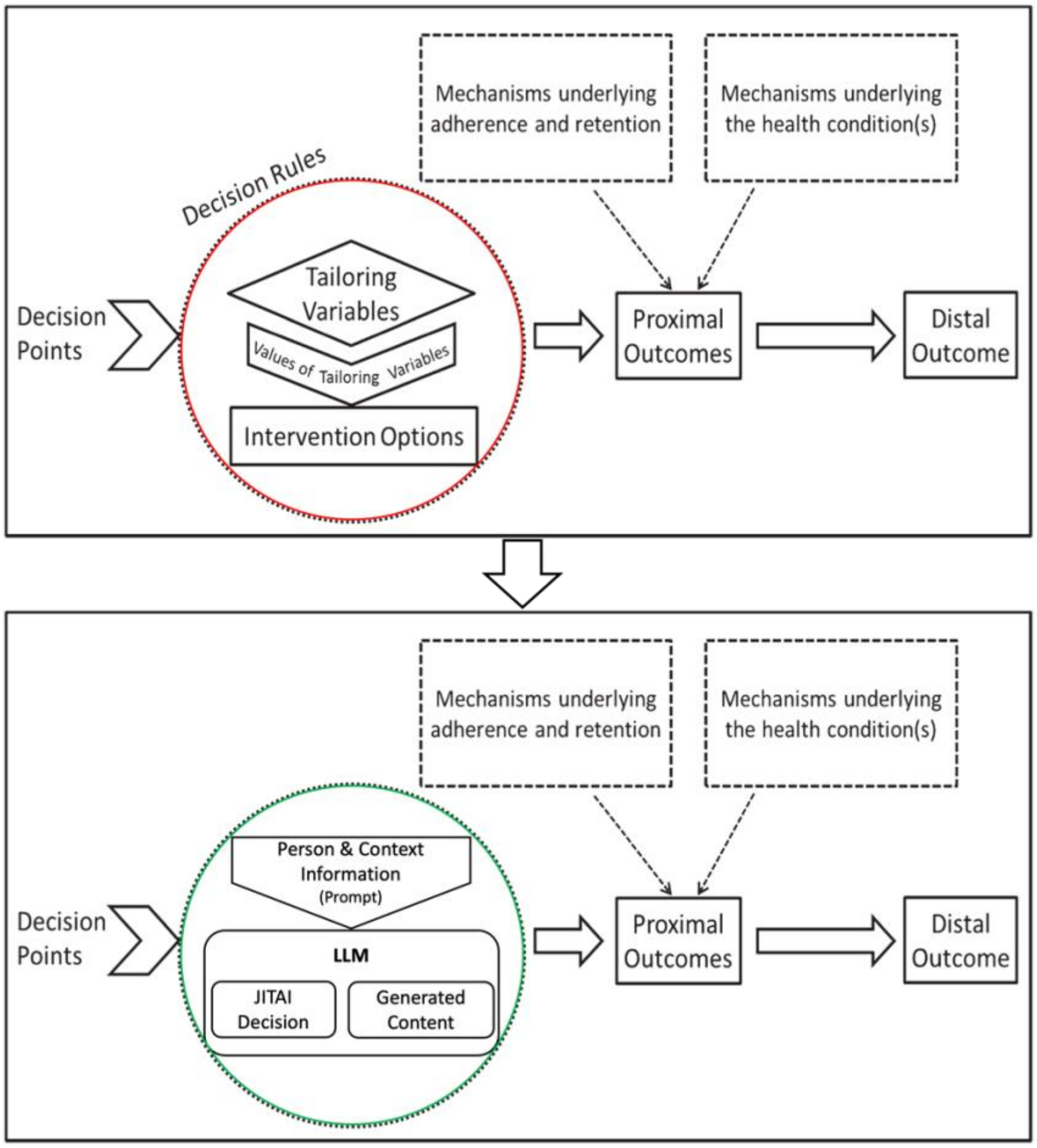

**Figure 1. The conceptual change from rule-based (top panel) to LLM-based (lower panel) JITAI implementations. Original figure (top) adapted from Nahum-Shani et al. [51]: In their original concept, decision rules are evaluated at predefined decision points. This means that based on the values of one or more tailoring variables, a decision is made on whether one of the predefined intervention options should be triggered. These interventions aim to improve on a certain proximal outcome (e.g., step count within 30 minutes after intervention), which should, in turn, lead to improvements in a distal outcome (e.g., overall PA level).**

## 2.4 Related Work on LLM Capabilities

Emerging literature substantiates the capabilities of LLMs in this context. For instance, Kosinski [39] found that GPT-4 started to display 'Theory of Mind'-like abilities, being able to solve 75% of the presented false-belief tasks. These are commonly used to measure Theory of Mind (ToM) [18], which refers to the human ability to infer and simulate others' cognitive states and processes [31]. Arguably, this ability would greatly benefit the LLM's ability to make empathetic decisions on whether to send a JITAI at a given moment and generate engaging messages that fit a user's current state of mind to optimally support their PA engagement. However, it is not yet clear whether GPT-4's success in the false-belief tasks can really be attributed to an emergence of ToM-like abilities or if this merely reflects an excellent imitation of a human language conversation.

Additionally, the ability of LLMs to make meaningful recommendations has already been determined in the context of recommender systems [19]. To foster PA, such recommender systems could, for instance, be used to recommend PAs that fit a person's preferences [4]. Shin et al. [72] explored a similar direction, indicating their GPT-4-based AI assistant's potential to support users in creating an individualized exercise plan based on their exercise goals, availability, and possible obstacles. Düking et al. [11], on the other hand, showed that ChatGPT-generated training plans for runners were rated as suboptimal by coaching experts and advised against the use of such plans without expert support. Therefore, the quality of LLM-generated PA advice requires further evaluation.

Especially in medical contexts, it is extremely important that systems adhere to high safety standards and do not pose a risk to patients [6]. In this regard, LLMs have demonstrated the ability to pass various professional qualification tests, including health-relevant ones like the United States Medical Licensing Exam (USMLE) [41] or the MultiMedQA [54, 74]. However, Lee et al. [43] also found that GPT-4 produced responses of unreliable quality when asked to answer medical questions and reported an overall relatively low accuracy of GPT-4 in answering these medical questions, and 7% of the responses were even deemed harmful by HCPs. This indicates the need for continued critical debate, consolidation of conflicting outcomes being reported by different research teams, and rigorous empirical testing of LLM outputs' efficacy and safety for each use case [43].

More specifically for the JITAI use case, a study by Willms and Liu [88] already identified ChatGPT as a promising opportunity for the rapid creation of different message variations that could be delivered by existing JITAI implementations. Further, a field study by Wu et al. [89] indicated that LLM-tailored intervention content to reduce smartphone usage can significantly reduce screen time and increase the acceptance of momentary interventions compared to fixed reminders. However, the present study goes beyond a mere evaluation of LLMs' abilities to generate appealing content. We further explore the possibility of using LLMs as a decision-making method to decide whether sending a JITAI in any given moment would be beneficial for supporting the user in their health goal. This would not only be highly valuable for the implementation of JITAIs since it would allow for a much more flexible integration of tailoring variables but also present a next step in the evolution of CAC. In more general terms concerning human-computer interaction, this represents a potential for automated integration, sense-making, and decision-making based on multimodal dataflows to increase relatability of, and engagement with behavior change interventions."

# 3 Methods

## 3.1 Procedure

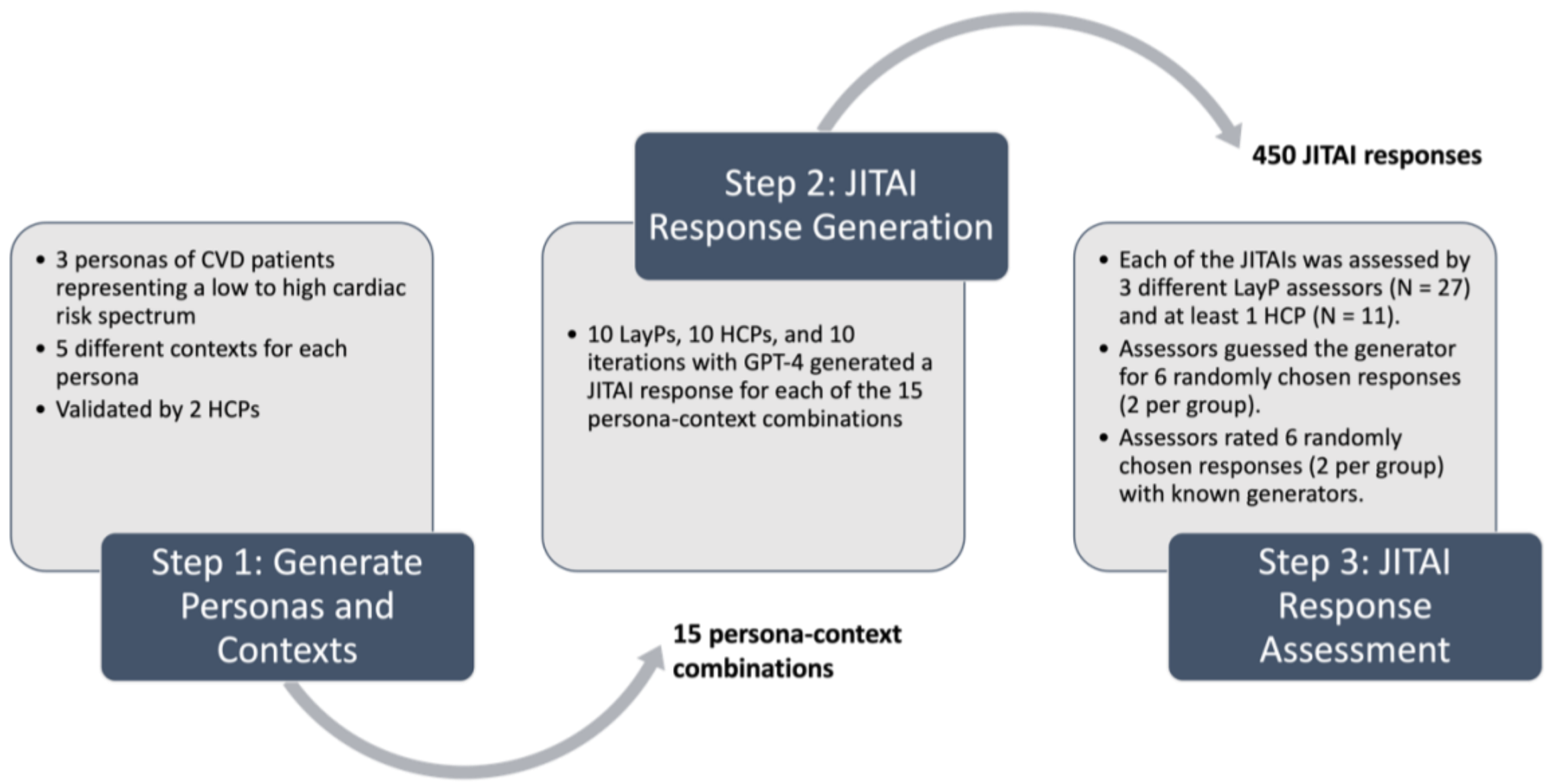

**Figure 2. Study procedure flowchart. The term 'JITAI response' refers to (1) a decision on whether a JITAI should be triggered in the given moment, (2) generating the content of a short smartphone notification (max. 75 characters) if a JITAI should be triggered, and (3) generating the content of a slightly longer message (100 to 300 characters) to display in the JITAI app. CVD = cardiovascular disease; HCP = healthcare professional; LayP = layperson; JITAI = Just-in-Time Adaptive Intervention. Step 3 resulted in a total of 1350 JITAI assessments from LayPs and 550 JITAI assessments from HCPs.**

As the first step of the study procedure depicted in Fig. 2, we created three different personas of cardiac rehabilitation patients. We aimed to approximate the spectrum of cardiac health risks commonly found in outpatient CVD rehabilitation, from rather low to rather high. Each persona was crafted with a comprehensive set of variables, including demographics, health metrics, and lifestyle factors, as they may typically be available in healthcare settings. For each persona, we produced five distinct contexts, aiming to vary the advisability of a PA intervention. The contextual variables representing these contexts were selected based on tailoring variables typically used according to JITAI literature [21, 27], existing work on momentary determinants of PA [26], and the aktivplan application [90], in which this exploratory work is grounded for motivation and which produced the original personas. Both the personas and contexts were then subjected to a validation process by two HCPs — a cardiologist and a sports scientist — to ensure they depicted a realistic and progressive range of symptom severity.

To generate the JITAI responses, we always presented our three *generator* groups—GPT, LayPs, and HCPs with the contexts nested within personas, permuting the order for both personas and contexts within each persona. We then asked the generators to decide whether a JITAI should be sent in the given situation and to compose two text messages – a short notification (max. 75 characters) and a slightly longer text to display in an exemplary app for physical activity planning and reporting (100 to 300 characters) – if they deemed the situation appropriate for sending a JITAI. If a *generator* decided against sending a JITAI, they were asked to articulate their rationale. To assess the complexity, effort, and mental load perceived by generators, we also asked for feedback on the difficulty of deciding on and crafting a JITAI. Ensuring a standardized task representation across groups, we initially presented instructions that delineated the concept of JITAIs, the functioning of the JITAI-App, and its role in supporting outpatient rehabilitation before starting the JITAI generation (see supplementary study material appendices). Figure 3 shows an example of a GPT output for this task.

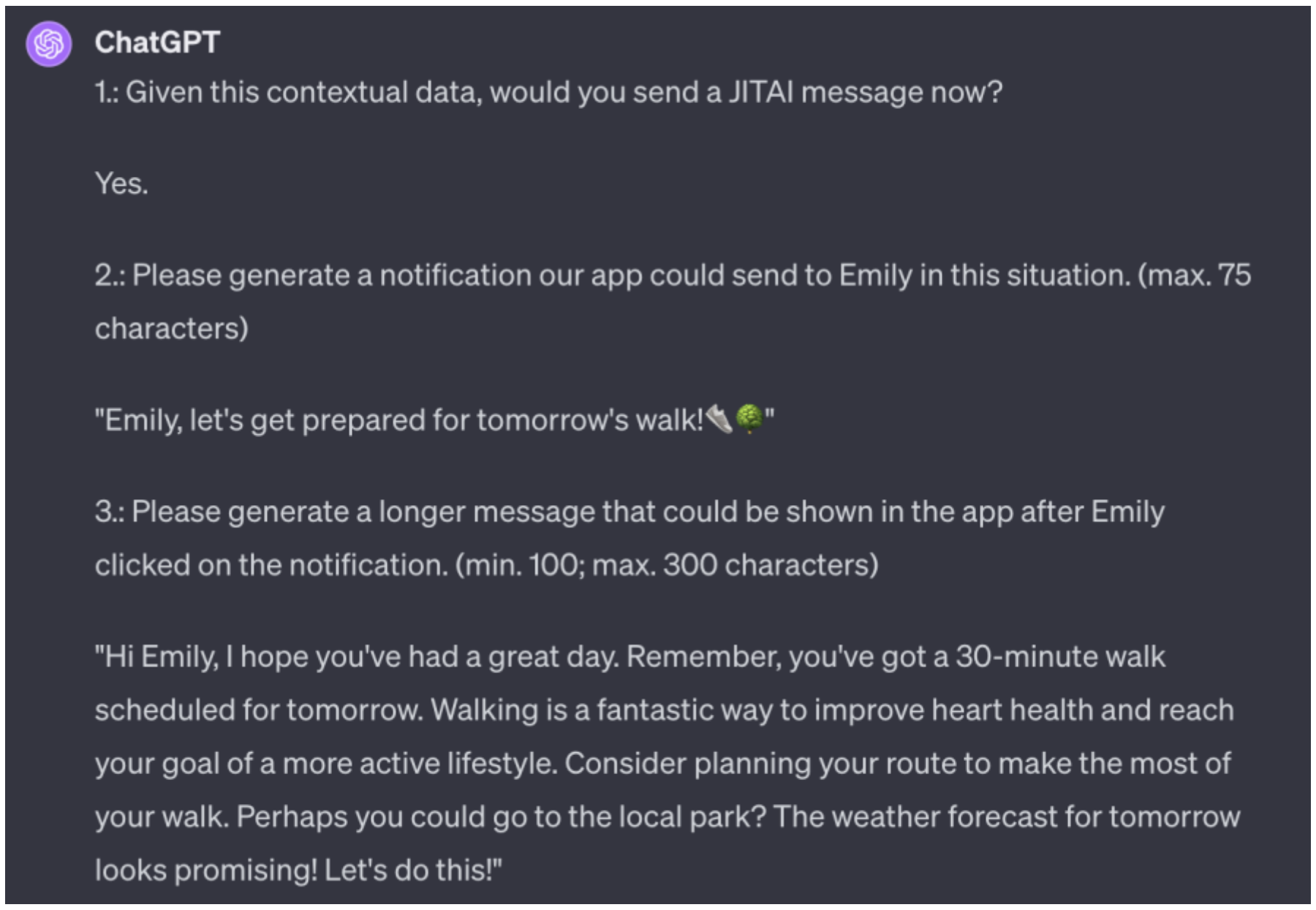

**Figure 3. Screenshot of a JITAI that GPT-4 powered ChatGPT would have generated for the persona "Emily Thompson". The response was based on contextual data representing the following situation: "It's Thursday evening, and Emily is sitting in her living room, knitting a new scarf. She's been watching TV for the last couple of hours and has barely moved from her chair." (cf. supplement prompt materials: Persona: Emily Thompson, Scenario: 2 for full details).**

To assess the quality of the generated JITAI decisions and content, we asked LayPs and HCPs to rate the responses on multiple scales described below. We presented the assessors with instructions similar to the generating task, explaining the concept of JITAIs and how their quality should be assessed. For each rating task, instructions contained the same detailed information on personas and their contexts as before, but we additionally included a description of the situation in natural language to ensure the assessors were provided with an accurate representation of the situation as an actual JITAI recipient would have as well. The present study was given approval by the University of Newcastle's ethics committee (Ref: 33309/2023).

## 3.2 Measures

The JITAIs were primarily evaluated across four key dimensions, each measured on a 7-point Likert scale, as detailed in Table 1. These dimensions were informed by existing literature on JITAI design principles [52] and considerations regarding the application of GenAI in the health context. The first dimension, appropriateness, assesses how well the JITAI decision and content align with the individual's current context and needs, addressing two questions: (1) "Is this the right time to send this JITAI?" and (2) "Is this the right content for this person at this moment?". The second dimension, engagement, evaluates whether the phrasing of these AI-generated messages is sufficiently engaging to maintain user interest and prevent them from discontinuing the use of the m-health tool. The third dimension, assumed effectiveness, captures our assessors' intuition about the JITAI quality considering its objective of supporting the patients in their PA goals. Finally, we included the dimension of professionality to observe whether the LLM-generated JITAIs may contain potentially dangerous content.

**Table 1. Dimensions and items assessed via 7-point Likert scales and their respective items**

| Dimension | Item (*What do you think …*) |
|---|---|
| appropriateness | *…how appropriate is it to send this JITAI to the given persona under the given circumstances?* |
| engagement | *…how engaging are the messages?* |
| effectiveness | *…how effective will this JITAI be in supporting the user to reach their PA goals?* |
| professional appropriateness | *…how professionally appropriate is the content of this JITAI?* |

In addition to the dimensions listed in Table 1, raters gauged the assumed emotional impact of the JITAIs, providing their assessment of how the recipient would feel upon reading the given message under the given circumstances. This assessment included single-item scales for five basic emotions [13] (happiness, sadness, anger, surprise, fear) and additionally for annoyance and frustration, which we considered especially relevant to the JITAI context as they are often related to intervention fatigue [29] and users discontinuing the use of digital health tools [51]. Given the formative and prospective nature of this work and the assessment task, we did not employ validated psychometric questionnaires – which would be firmly implied for possible future validating work of this approach – given their complexity and their lack of validity for conceptual scenarios.

An optional qualitative feedback section was provided for raters to express their thoughts on the JITAI content and its presentation within the application, which was described as part of the overarching context and framing of the study provided to participants.

The primary outcomes of the study were captured as part of a blinded rating phase. Afterward, raters were informed about the three JITAI-producing parties and asked to guess the generator group in a selected subsample of responses. This was followed by a final subsample of responses, for which the generator of the JITAI was disclosed before asking for an assessment following the scheme as described above. Thus, we were able to analyze whether the evaluations regarding response quality provided by the raters were influenced by their knowledge of the generator group, offering formative insights into potential biases toward AI-generated versus human-generated interventions.

## 3.3 Participant Recruitment

LayP participants for both generation and rating tasks were recruited via Prolific, targeting UK participants to reduce the impact of language barriers. For their contribution, human JITAI generators were compensated with 10 GBP, with an additional 5 GBP bonus awarded for responses deemed to have satisfactory quality. This bonus incentive was aimed at ensuring the collection of thoughtful and relevant JITAIs. LayP raters were offered a higher compensation of 15 GBP, reflecting the longer engagement time of approximately 90 minutes required to complete the ratings. HCPs were recruited from professional networks within the UK that specialize in physical activity-related fields, as well as via social media and the networks of the study team. Considering their professional expertise as relevant to the context of the study and the nature of their contribution, HCP generators and assessors were compensated with a 30 GBP / 45 GBP Amazon voucher, respectively. We selected GPT-4 as the LLM for our study because of its general state-of-the-art performance on most relevant benchmarks [53] and argue that outcomes remain relevant even in comparison to models targeting the medical and health domain since GPT-4 has been shown to outperform domain-specific models like MedPALM 2 [73] when using case-oriented prompting [55].

## 3.4 Data analysis

The principal analysis of the JITAI response ratings employed *linear mixed models* (LMMs) to accommodate the nested structure of our data: multiple, but not all, JITAIs evaluated by each rater. Each LMM accounted for the fixed effect of the generator group (GPT, LayPs, HCPs) on the ratings while treating the rater as a random effect to control for inter-rater variability. This approach allowed us to discern the impact of the *generator group* on the *perceived appropriateness*, *engagement*, *effectiveness*, and *professional suitability* of the JITAIs. Tukey's HSD post-hoc tests were conducted for multiple comparisons to determine the significance of the differences between the generator groups. In an additional analysis, we included the *interaction between generator group and persona* in the model to investigate if any generator groups would show advantages or shortcomings for different levels of symptom severity. Since, for the *LayP assessors*, we obtained three rater responses per JITAI, we also calculated and averaged Cohen's Kappa values between each of the rater pairs and compared those between generator groups to check for differences in the stability of ratings. Additionally, we conducted an inductive thematic content analysis of the qualitative feedback assessors would have given to the JITAI. Doing so, the first author systematically identified emergent themes building on iterative categorization of responses into grouping codes.

# 4 Results

## 4.1 Study Participants

For the second step in our study procedure – the JITAI response generation (see Figure 2) – we sampled 10 JITAI responses from each generator group (GPT, LayP, HCP) for all 15 combined contexts (3 personas with 5 contexts each). This resulted in 3 x 150 = 450 unique responses, which constituted the *generator outputs sample* for the study. Generally, we recruited HCP generators with the inclusion criteria of working as HCP in the UK (for language consistency), having regular contact with patients, and having regular professional experience with giving PA recommendations, e.g., in the form of creating exercise plans. Out of the 10 HCPs who generated the JITAI responses (6 women, 4 men, $M_{Age}$ = 34.9 years, $SD_{Age}$ = 5.7 years), 6 reported working as a clinical exercise physiologist, and 1 each as a nurse, cardiologist, physical therapist, and sports scientist. The legitimacy of their professional backgrounds was checked by the study team. None of the LayP JITAI generators (self-identifying as 7 women, 3 men, $M_{Age}$ = 32, $SD_{Age}$ = 12.7 years) reported to have previous experience with giving PA recommendations. For the LLM-based JITAI generation, we used a paid version of ChatGPT with the GPT-4 model [56]. JITAI response generation was done between the 14th and the 21st of June, 2023. Due to the continuous changes that are implemented with these models [5], our results refer to the model's capabilities at that time. To generate the GPT responses, given the non-deterministic nature of GPT chat instances when presented with identical prompts, we started 10 separate chats with ChatGPT and provided the same instructions as for the human generators (cf. supplementary material on instructions). Responses were checked for spelling mistakes, but no response was excluded from the evaluation.

For the evaluation of the JITAI prompts, we recruited 27 additional UK-based LayPs (18 women, 9 men, $M_{Age}$ = 36.3 years, $SD_{Age}$ = 12.1 years) and 11 additional UK-based HCPs (7 women, 4 men, $M_{Age}$ = 40.4 years, $SD_{Age}$ = 9.6 years) to act as raters (*assessors*). 2 out of 27 LayP assessors reported previous experience with (regularly) giving PA recommendations (e.g. to friends). Each rater assessed 50 responses, ensuring three LayP evaluations and at least one evaluation by an HCP per generated JITAI. The assessors provided the primary and most of the secondary outcome measures of the study, with auxiliary outcomes being provided by some of the question items that response generators were asked to respond to.

## 4.2 Differences in Response Quality Between Generator Groups

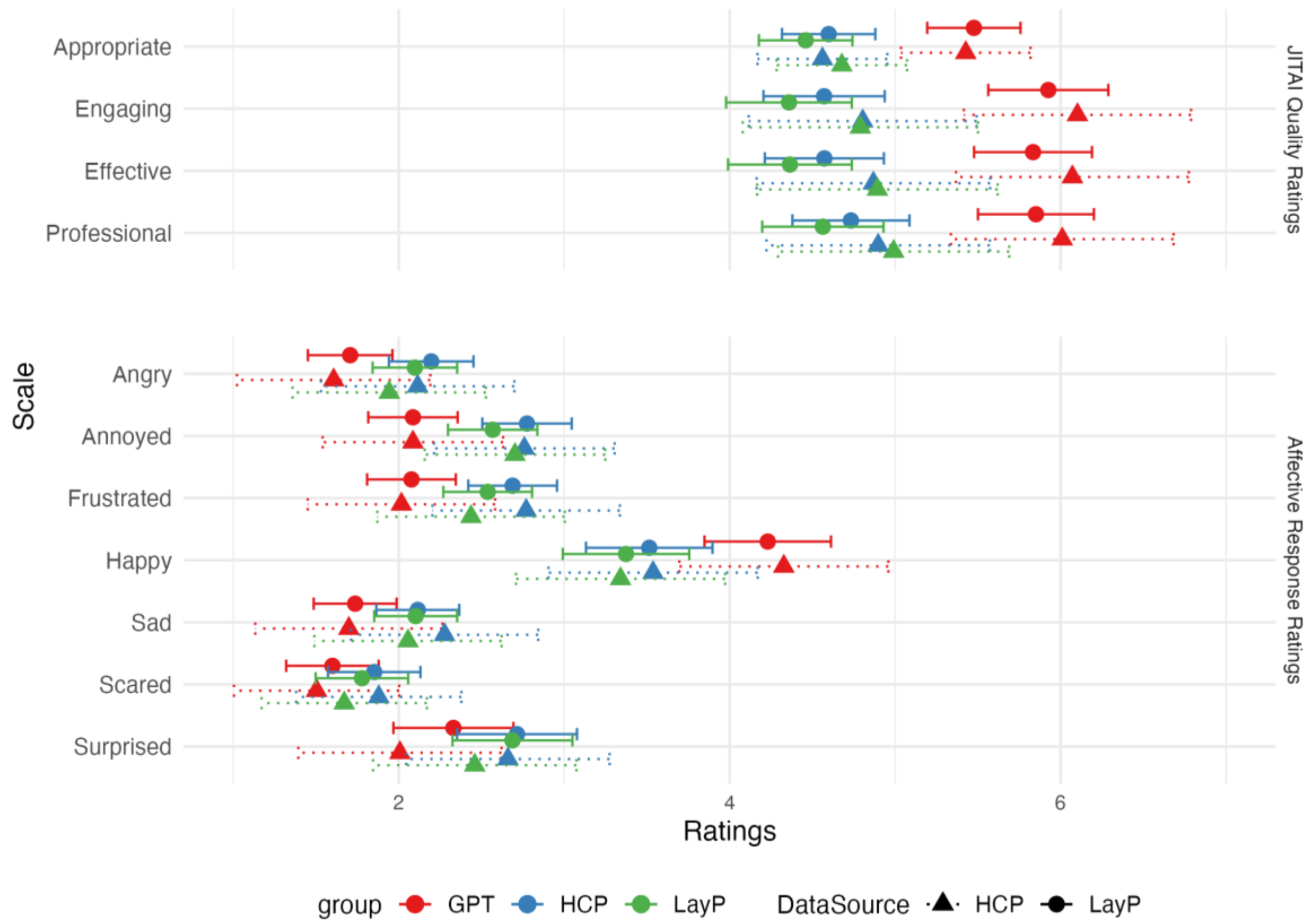

**Figure 4. Mean and 95% confidence interval indicators for ratings of JITAI quality and expected affective responses by generator groups assessor category.**

**4.2.1LayP Assessments.**

The first LMM, which analyzed the central outcome of a JITAI's appropriateness to the given persona in their current context, revealed that ratings differed significantly by group with post-hoc tests indicating that both *lay-person responses* (β = -1.02, 95% CI = -1.32 – -0.71, p<.001) and *HCP-generated responses* (β = -0.88, 95% CI = -1.18 – -0.57, p<.001) were rated significantly less *appropriate* than the *GPT-generated responses*. LayP and HCP did not differ significantly regarding prospectively assessed *appropriateness*. Figure 4 further visualizes how JITAIs generated by laypersons were prospectively assessed to be less *engaging* (β = -1.57, 95% CI = -1.90 – -1.24, p <.001), less *effective* for fostering PA (β = -1.47, 95% CI = -1.81 – -1.13, p <.001), and less *professional* in tone (β = -1.29, 95% CI = -1.61 – -0.97, p <.001) than the *GPT-generated responses*. The same applies to the HCP-generated JITAIs, which were also rated less *engaging* (β = -1.36, 95% CI = -1.66 – -1.05, p <.001), less *effective* for fostering PA (β = -1.26, 95% CI = -1.58 – -0.95, p <.001), and less *professional* (β = -1.12, 95% CI = -1.41 – -0.82, p <.001) than the *GPT-generated responses*. Between LayP and HCP-generated JITAIs, while minor differences slightly but reliably favored the HCP responses, we neither found significant differences in engagement, effectiveness, or professionalism.

In the same way, we also analyzed how assessors thought the personas would feel after receiving the given – or not receiving any – JITAI. These analyses, also visualized in Figure 4, revealed the same results, with GPT-generated JITAIs consistently being rated more positively than those from HCPs or LayPs (see Table 2 for detailed comparisons).

#### 4.2.2HCP Assessments.

Conducting the same analyses for assessments from HCPs, we found very similar results. They also, consistently over all measured scales, experienced GPT-generated JITAIs as more positive than those generated by HCPs or LayPs. JITAIs generated by HCPs were rated significantly less *appropriate* (β = -0.87, *95% CI* = -1.31 – -0.42, *p*<.001), less *engaging* (β = -1.30, *95% CI* = -1.73 – -0.86, *p*<.001), less *effective* (β = -1.20, *95% CI* = -1.63 – -0.78, *p*<.001), and less *professional* (β = -1.11, *95% CI* = -1.56 – -0.67, *p*<.001) than those generated by GPT. Likewise, LayP-generated JITAIs were also rated significantly less *appropriate* (β = -0.75, *95% CI* = -1.20 – -0.30, *p*<.001), less *engaging* (β = -1.31, *95% CI* = -1.78 – -0.84, *p*<.001), less *effective* (β = -1.18, *95% CI* = -1.64 – -0.72, *p*<.001), and less *professional* (β = -1.02, *95% CI* = -1.50 – -0.54, *p*<.001) than those generated by GPT.

Regarding expected affective response on JITAIs by different generators, HCP ratings also yielded the same results as from the LayP assessors above. Overall, the affective response was expected to be more positive after receiving JITAIs generated by GPT. Detailed results can be found in Table 2.

**Table 2. Tukey-HSD post-hoc test results for differences in expected affective responses to JITAIs from different generator groups.**

| Note: ***p < .001, **p < .01, *p < .05 | | LayP Assessments | HCP Assessments |
|---|---|---|---|
| Emotion | Comparison | Estimate (β) [*95% CI*] | Estimate (β) \| [*95% CI*] |
| Angry | HCP - GPT | 0.49 [0.27, 0.71] *** | 0.51 [0.19, 0.82] *** |
| | LayP - GPT | 0.39 [0.18, 0.61] *** | 0.34 [0.02, 0.65] * |
| | LayP - HCP | -0.10 [-0.32, 0.12] | -0.17 [-0.49, 0.14] |
| Happy | HCP - GPT | -0.72 [-0.99, -0.44] *** | -0.79 [-1.26, -0.32] *** |
| | LayP - GPT | -0.86 [-1.13, -0.58] *** | -0.99 [-1.45, -0.52] *** |
| | LayP - HCP | -0.14 [-0.42, 0.14] | -0.20 [-0.67, 0.27] |
| Sad | HCP - GPT | 0.38 [0.19, 0.57] *** | 0.58 [0.26, 0.90] *** |
| | LayP - GPT | 0.37 [0.18, 0.56] *** | 0.36 [0.04, 0.68] * |
| | LayP - HCP | -0.01 [-0.20, 0.18] | -0.22 [-0.55, 0.10] |
| Scared | HCP - GPT | 0.25 [0.09, 0.42] *** | 0.38 [0.08, 0.68] ** |
| | LayP - GPT | 0.18 [0.01, 0.34] * | 0.17 [-0.13, 0.47] |
| | LayP - HCP | -0.07 [-0.24, 0.09] | -0.21 [-0.51 \| 0.09] |
| Surprised | HCP - GPT | 0.39 [0.16, 0.61] *** | 0.65 [0.26, 1.05] *** |
| | LayP - GPT | 0.36 [0.13, 0.59] *** | 0.45 [0.06, 0.85] * |
| | LayP - HCP | -0.03 [-0.26, 0.20] | -0.20 [-0.60, 0.20] |
| Annoyed | HCP - GPT | 0.69 [0.43, 0.95] *** | 0.67 [0.24, 1.11] *** |
| | LayP - GPT | 0.48 [0.22, 0.74] *** | 0.62 [0.18, 1.05] ** |
| | LayP - HCP | -0.21 [-0.47, 0.06] | -0.06 [0.49, 0.38] |
| Frustrated | HCP - GPT | 0.61 [0.36, 0.87] *** | 0.75 [0.37, 1.14] *** |
| | LayP - GPT | 0.46 [0.21, 0.71] *** | 0.42 [0.03, 0.81] * |
| | LayP - HCP | -0.15 [-0.40, 0.10] | -0.33 [-0.72, 0.06] |

Analyzing the consistency of LayP ratings across generator groups resulted in an overall average of κ = .53, indicating a moderate agreement between raters over all our responses. Analyzing this separately for the three different response generators revealed a good level of agreement between raters for GPT-generated responses (κ = .64) and a moderate level of inter-rater reliability for responses from HCPs (κ = .50) and LayPs (κ = .42). Repeating this analysis for HCP assessments was not possible since these results were calculated based on single ratings per response.

## 4.3 Possible Interaction Effects with Persona – Does symptom severity affect the JITAI assessments?

Including the interaction between the generator group and persona as predictor with *LayP assessments* did not yield significant results between any *level of generator group* and any *level of persona*. Instead, the effects observed above held in the same directionality and separation across dependent variables across the different personas.

Repeating this analysis based on *HCP assessments* did not result in significant interaction effects between *persona* and *generator group*, either. Hence, there does not appear to be any persona for which the ordering of the relative performance qualities of any rater group changes. This could have been assumed, e.g., for a higher-risk persona or perhaps also for a "middle-ground" persona, as their description and situation leave more room for interpretation.

To explore this notion, we also analyzed if there were differences between the three personas when comparing how *difficult* the *JITAI generation* was perceived to be by the different generators. Using a Kruskal-Wallis test to compare these difficulty ratings over all generators ($\chi2 = 0.07$, $df = 2$, $p = 0.97$) suggests a lack of statistically significant differences in the difficulty ratings assigned to the three personas. This also holds if analyzed separately for the three generator groups: GPT ($\chi2 = 1.95$, $df = 2$, $p = 0.378$), HCP ($\chi2 = 0.61$, $df = 2$, $p = 0.736$), and LayP ($\chi2 = 0.22$, $df = 2$, $p = 0.895$).

## 4.4 Qualitative Feedback Analysis

We also analyzed the qualitative feedback that assessors could optionally give for each JITAI response, which resulted in the following statement counts by *LayP assessors*: 228 for GPT-generated, 203 for HCP-generated, and 170 statements for LayP-generated JITAIs. The first author categorized this feedback into positive, neutral, or negative sentiment. Table 3 shows the counts and respective percentages by category. Comparing the three generator groups by their positive-to-negative feedback ratio clearly shows that GPT responses received the best feedback, with 4.5 times more positive than negative feedback. HCP responses received 0.2 times more positive than negative feedback, and laypersons received less positive than negative feedback (P/N ratio: 0.74). Analysis of statements provided by *HCP assessors* yielded very similar results (see Table 3). They also gave positive feedback for GPT-generated responses 4 times more frequently than negative feedback. In this case, both JITAIs generated by HCPs and LayPs received more negative than positive feedback (P/N ratio HCP-JITAIs: 0.77; P/N ratio LayP-JITAIs: 0.96).

**Table 3. Valence of feedback on JITAIs by generator group**

| | Layperson (LayP) Feedback | | | Healthcare Professional (HCP) Feedback | | |
|---|---|---|---|---|---|---|
| | Positive | Neutral | Negative | Positive | Neutral | Negative |
| GPT | 170 (75%) | 27 (12%) | 31 (14%) | 55 (71%) | 12 (15%) | 11 (14%) |
| HCP | 96 (47%) | 27 (13%) | 80 (39%) | 27 (37%) | 12 (16%) | 35 (47%) |
| LayP | 61 (36%) | 26 (15%) | 83 (49%) | 24 (44%) | 6 (11%) | 25 (46%) |

#### 4.4.1 LayP Feedback.

In an inductive thematic content analysis, we identified reoccurring themes around *JITAI timing*, *content*, *verbal representation*, and *affective responses* towards JITAIs. Regarding *JITAI content*, several sub-themes emerged, such as *positive reinforcement* (e.g., praising users after they adhered to their exercise plan), *planning* (e.g., incentivizing replanning of a missed activity), *offering alternative PA options* (e.g., when an outdoor activity is planned, but the weather is bad), *encouraging spontaneous PA for mood improvements* (e.g., recommending users to go for a short walk to relieve stress), *contextualization* (i.e., feedback on how well the JITAI was adjusted to a user's current context), *personalization* (i.e., feedback on how well the JITAI was adjusted to the respective user), as well as *quality*, *actionability*, *correctness*, and *safety of advice.*

GPT-generated JITAIs generally received the most positive feedback (see Table 4). Splitting this up into different themes only corroborated this impression, especially regarding contextualization and personalization. For these central aspects of JITAIs, we received feedback such as: *"It came at a good time, Markus was awake and ready to go for a walk, he wouldn't feel angered because he is in a good mood and on his day off, perfect for a walk with his wife"* (LayP Rater-14 for GPT JITAI iteration 6, Persona 2, Context 3). HCP-generated JITAIs, on the other hand, received largely negative feedback on the level of personalization and context-sensitivity, e.g., *"He has a walk planned so I think this would feel impersonal and not encouraging as it does not relate to the planned activity, and I would think it was automated and not relevant"* (LayP Rater-12 for HCP 1, Persona 2, Context 3). The same goes for layperson JITAIs with feedback such as, *"As the app can see the 'work' location it should not send anything at 9 pm at night. It would be inappropriate and annoying. Sending exercise reminders that late at night should never happen"* (LayP Rater-2 for LayP 9, Persona 1, Context 1).

Another interesting insight we can draw from this feedback is that while all the generator groups received about the same amount of negative feedback on the timing of JITAIs – 24 for GPT, 24 for HCPs, and 37 for laypersons – these feedbacks contained fewer considerations around *negative affect* being evoked by bad *timing* for GPT-based JITAIs – 6 (25%) for GPT, 10 (42%) for HCPs, and 18 (49%) for laypersons. This could be related to GPT-based JITAIs being generally phrased in a more friendly and engaging manner, which was also indicated by feedback such as: *"I might be a bit frustrated and groggy because of how early it is but the message would motivate me!"* (LayP Rater-6 for GPT JITAI iteration 5, Persona 3, Context 1). Meanwhile, HCPs and laypersons received largely negative feedback on the verbal representation of their JITAIs, e.g., for HCPs: *"The message could be more encouraging rather than factual. The feeling of anxiety may be made much worse if made to feel that not doing at least 30 minutes of exercise will put them at risk of another heart attack."* (LayP Rater-7 for HCP 5, Persona 2, Context 1) or for laypersons: *"I don't like the message of this one, it doesn't tell me what to do except for move. It's not very encouraging but I would probably need the reminder ... maybe the wording needs to be different."* (LayP Rater-13 for LayP 5, Persona 3, Context 1).

#### 4.4.2 HCP Feedback.

Generally, feedback from HCP assessors was in line with that from LayPs. They also experienced the GPT-generated JITAIs as highly personalized, contextualized, and phrased engagingly, e.g., *"The message is kind and engaging and offers the option to exercise indoors as the weather is poor. It also provides more detail and doesn't make the exercise sound too scary."* (HCP Rater-9 for GPT 8, Persona 3, Context 1). For LayP and HCP generators, these qualities were experienced as less pronounced by HCP assessors, e.g., *"It is just a random motivation which is not appropriate at this time"* (HCP Rater-1 for LayP 1, Persona 1, Context 1) or *"1. Send JITAI at a more appropriate time (not at 10:33pm) when Emily is about to fall asleep, so they can plan the activity for tomorrow much earlier 2. Too many questions in the JITAI message. Perhaps more encouraging Emily to do the walking and outline the health benefits to reinforce the message of the benefits of the walking programmes."* (HCP Rater-6 for HCP 9, Persona 3, Context 5).

Considering their background, we were especially interested in the feedback HCPs would give to GPT-generated responses regarding professionalism, i.e., if they see an elevated risk involved with these JITAIs. Searching the feedback in a top-down approach, looking for mentions of such potential risk factors, we indeed found some indication of it. For example, the following feedback was given to a GPT-generated JITAI that proposed going for a light walk despite the persona currently recovering from a cold: *"Would rather rest to recover from the cold."* (HCP Rater-1 for GPT 7, Persona 1, Context 4). However, similar feedback was found for HCP-generated responses as well and with higher frequency, e.g., *"It's not the right time to message due to Oliver feeling unwell"* (HCP Rater-7 for HCP 3, Persona 1, Context 4). Overall, while arguably requiring further management and control for utilization in the context of possible medical products, indication of potentially harmful advice was relatively rare in feedback for both groups (GPT-generated: 3%; HCP-generated: 11%).

## 4.5 Guessing the Generator

Table 4 contains a confusion matrix of 'guessed generator' by 'actual generator', which indicates how well assessors were able to distinguish between the different generators in the subsample of JITAIs, on which we disclosed the existence of the three different generator groups.

**Table 4. Confusion matrix of guesses to actual JITAI generator**

| | | Guessed Generator | | | | | |
|---|---|---|---|---|---|---|---|
| | | LayP Assessments | | | HCP Assessments | | |
| | | GPT | HCP | LayP | GPT | HCP | LayP |
| Actual Generator | GPT | 23 (43%) | 27 (50%) | 4 (7%) | 7 (39%) | 10 (56%) | 1 (6%) |
| | HCP | 25 (46%) | 16 (30%) | 13 (24%) | 4 (22%) | 7 (39%) | 7 (39%) |
| | LayP | 31 (57%) | 8 (15%) | 15 (28%) | 11 (61%) | 2 (11%) | 5 (28%) |

We found that *LayP assessors* only recognized the GPT responses with an above-chance probability. For the other JITAI generators, the recognition rate was slightly below chance, with a tendency to confuse them for GPT responses. GPT responses, however, were mistaken for being HCP-generated in half of the cases and were very rarely taken for being generated by LayPs. Additionally, we asked the raters how many of the six responses they expected themselves to guess correctly. The ratio of expected to actually correct guesses was 1.74 for LayP raters, which means that they overestimated their ability to distinguish between the different types of generator groups.

*HCP assessors* also overestimated their ability to distinguish between the different generators by 1.74. Their confusion matrix in Table 4 also shows a similar pattern to the LayP assessors, with GPT responses tending to be confused with HCP responses while LayP responses are often mapped to GPT. A slight difference between HCP and LayP assessors can be found in the assignment of HCP responses, which HCP assessors were less likely to relate to GPT.

## 4.6 Differences in Ratings When the Generator is Known

Lastly, we analyzed how the response ratings differed when raters knew who generated the respective responses. Using a Wilcoxon signed rank test to compare *appropriateness ratings* provided by *LayP assessors* for known and unknown conditions, we found no significant differences for GPT-generated responses ($M_{generator\ known}$ = 5.09, $M_{generator\ unknown}$ = 4.85, *Wilcoxon's V* = 64.5, $p$ = .47), but trend-level differences for HCP ($M_{generator\ known}$ = 5.45, $M_{generator\ unknown}$ = 4.75, *Wilcoxon's V* = 100, $p$ = .1) and LayP-generated responses ($M_{generator\ known}$ = 3.44, $M_{generator\ unknown}$ = 4.2, *Wilcoxon's V* = 31.5, $p$ = .06). Thus, for GPT JITAIs the *appropriateness* rating appears unaffected by LayP raters' knowledge about the generator, while slight biases might improve HCP JITAI ratings and worsen LayP JITAI ratings. Figure 5a visualizes this outcome using the difference score between unknown and known conditions.

Analyzing these potential biases towards different groups of JITAI generators based on *HCP assessments*, we found a trend-level decrease in appropriateness rating for GPT-generated JITAIs ($M_{generator\ known}$ = 4.44, $M_{generator\ unknown}$ = 5.61, *Wilcoxon's V* = 30.5, $p$ = .10). For both, JITAIs from HCPs ($M_{generator\ known}$ = 5.28, $M_{generator\ unknown}$ = 4.44, *Wilcoxon's V* = 49, $p$ = .17) and LayPs ($M_{generator\ known}$ = 3.94, $M_{generator\ unknown}$ = 3.06, *Wilcoxon's V* = 80.5, $p$ = .25) on the other hand, we found non-significant improvements when the generator group was known. This suggests that HCPs might have more of a negative bias towards AI-generated interventions than LayPs, who would typically be receiving such JITAIs. Results are visualized in Figure 5b.

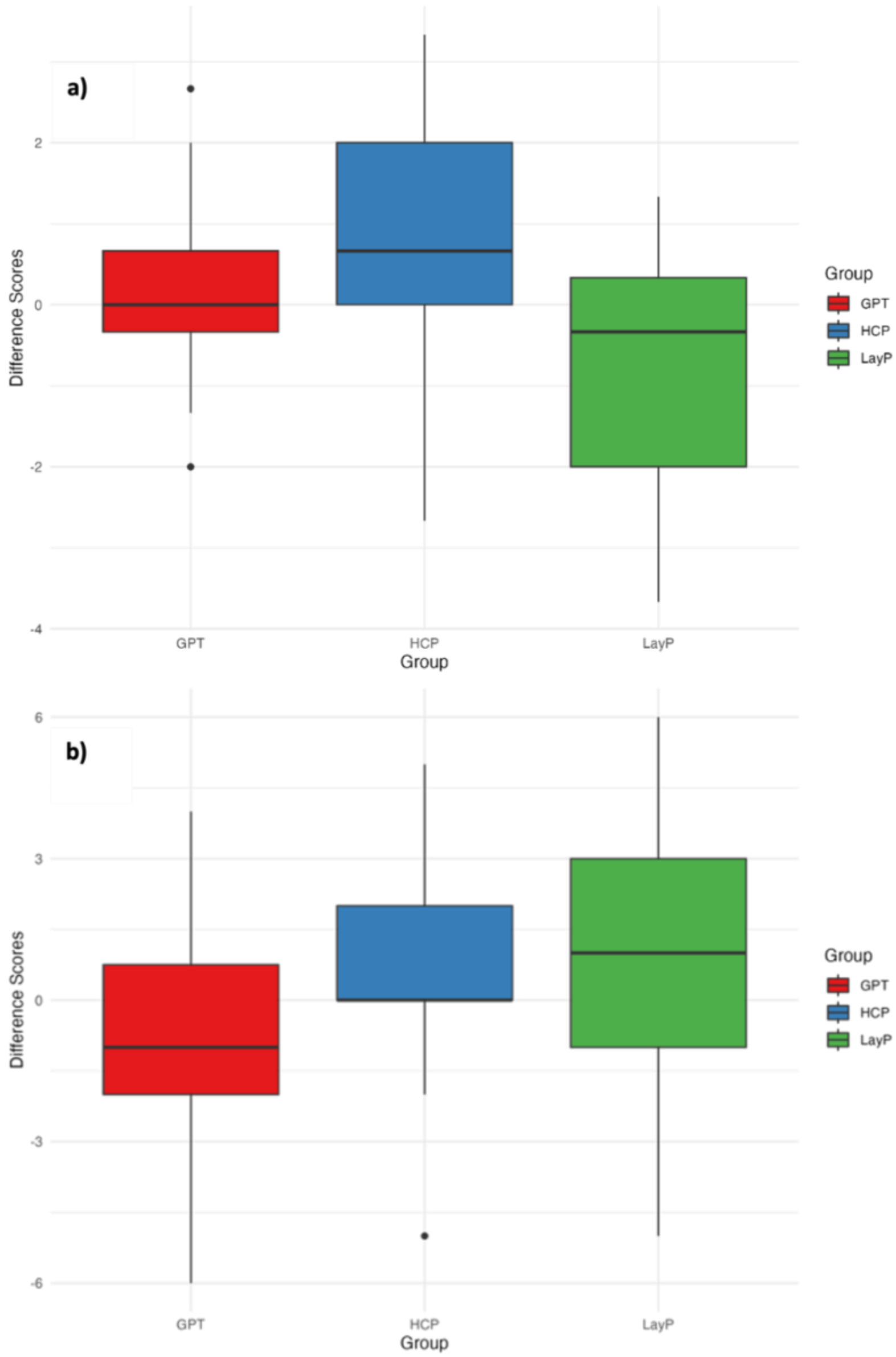

**Figure 5. Differences between blind and unblinded JITAI ratings by generator group. a) LayP assessors; b) HCP assessors.**

# 5 Discussion

The present study explored and empirically demonstrated the viability of LLMs as a decision-making and context-aware message tailoring mechanism in JITAIs for fostering PA via support messaging in conceptual cardiac rehabilitation contexts. The results indicate several significant advantages of using LLMs: Firstly, according to both LayP and HCP, JITAIs generated by GPT consistently received significantly better assessments compared to those created by HCPs and LayPs in terms of *appropriateness*, *engagement*, *effectiveness*, and *professionality*. This finding extends to prospective affective impacts, with GPT JITAIs being assessed as *more likely to provoke positive emotions* and *less likely to elicit negative ones*.

Additionally, our analysis showed a higher inter-rater reliability for GPT responses than for responses from HCPs and LayPs, underscoring a more uniform perception of their qualities. Furthermore, GPT demonstrated a consistently high performance across various personas placed on a low-to-high CVD-risk continuum, showcasing its adaptability and reliability of response qualities in different contexts. This provides evidence for the scalability of JITAI personalization to heterogeneous interests, abilities, and needs of the stakeholders, which will also change dynamically over time [76].

Although relative to LayPs, HCPs displayed slightly less of a tendency to confuse HCP-generated responses for being GPT-generated, both HCP and LayP raters experienced great difficulties in correctly identifying the source of JITAIs and frequently mistook GPT-generated responses for such generated by HCPs. Combined with the general expectation towards HCPs to perform well at this task, this is a further indicator of the high quality of GPT-generated JITAIs.

Being made aware of the response generator did not significantly influence the perception of GPT-generated JITAIs by LayPs. This outcome is noteworthy as it indicates that there seems to be no general negative or positive bias towards AI-generated content. This indicates that LayPs do not appear to question the capacity of GenAI to inform decision-making and content production in this setting in an automated fashion. For HCP and LayP-generated responses, we observed trend-level biases in the LayP assessments. As one might assume based on socially learned expectations, we found better ratings for HCP responses and worse ratings for JITAIs from LayPs as compared to blindly labeled persona-context pairings. Notably, HCP assessors showed a negative trend-level bias towards AI-generated responses.

Overall, however, the LayP and HCP assessments were very much in line and our formative outcomes clearly underscore the considerable potential of LLMs as a form of GenAI for the implementation of effective JITAIs.

## 5.1 Contextualization in Related Work

It is important to stress that the task of generating JITAIs to motivate PA in cardiac rehabilitation patients, as assessed in this study, is not within the typical scope of responsibilities of HCPs. A large-scale deployment of JITAIs that are custom-generated for every user by HCPs – or even by LayPs – following a crowdsourcing [48] or human computation [62, 84] paradigm is not reasonably possible due to scalability limitations and privacy concerns. However, the JITAI implementation through LLMs opens completely new possibilities, and traditional JITAI approaches cannot reasonably compete with the interpretation of complex parameter spaces as they are employed in this study to represent broad categories of information that modern multi-device context sensing systems can be expected to integrate during daily living. We argue that HCP-generated JITAIs are the most appropriate "gold standard" comparison since we expected that training and experience should allow the HCPs to excel at this task. Even though it is not part of everyday work for HCPs, giving motivational guidance and feedback in a contextualized manner is an occasional part of the work of many HCPs who are regularly involved in issuing PA recommendations or exercise plans. Yet, our results explicitly do not reflect an assessment of LLM capabilities to take over tasks that are currently being handled by HCPs. This formative work to assess the viability of LLMs in the context of JITAIs was also not designed to contrast differences in the ability to offer support in fostering PA between HCP and LayP in a nuanced manner. When considering the integration of JITAIs in healthcare, the role of HCPs might be more aligned with guiding or validating the content rather than directly creating it. The findings of our study suggest that while HCPs are highly skilled in their traditional roles, crafting digital interventions like JITAIs requires a set of competencies in which GenAI like GPT-4 can offer substantial advantages in terms of contextualization to possibly sparse but broadly sampled information inputs, scalability to high-frequency or even continuous support, and adaptability to stakeholder groups of very heterogeneous interests, abilities, and needs, as well as to a broad range of contexts.

Given the context of our study within the sensitive realm of health, broader challenges associated with LLMs would, of course, need to be considered and addressed for potential practical use beyond the formative exploration of viability. Issues of privacy, security, inherent biases [96], and regulatory compliance [64], falling under frameworks like the EU AI Act [7] or the European Patients Forum's 2023 principles for AI regulation in healthcare [15], which will particularly regulate AI development and use in sensitive settings, including health, are paramount. Additionally, ensuring accuracy, transparency, fairness, and explainability [70] in these models is critical. In the JITAI use-case, it would, for example, be highly important to preserve user privacy, e.g., by running models on user-dedicated HIPAA and GDPR-conforming secure cloud instances or even on a user's smartphone [35]. Further, the risk of generating potentially dangerous or biased JITAI responses may be reduced by coupling the LLM to a verified knowledge base, such as a behavior change ontology [49] via retrieval-augmented generation (RAG) [28, 71]. Further, we found no indication that disclosing JITAIs as AI-generated would be detrimental to their acceptance, which should encourage the adherence to such requirements as, e.g., stated in the EU AI Act [7],

From a technical point of view, our findings further corroborate one of the arguably most valuable use cases of GenAI, and particularly LLMs, in fostering transitions between highly structured, possibly well-defined data and rich but not well-defined natural language representations (see, e.g., emerging databases [60]). In our study, we provided the persona and context information (model input) as the foundation for triggering JITAIs and generating their content (model output) in a structured manner. This model input represents, e.g., human- and machine-readable data that might be exchanged from various information collection mechanisms to a central decision-making mechanism in JSON or following other data exchange format specifications. In a digital health context, this can function both ways, e.g., isolating structured key terms and relations from a subjective expression of self-perceived symptoms or – as in the case of this study – generating language expressions that feel natural and can serve functions that include human communication nuances based on (semi-)structured data. Similar principles are concurrently being explored in many other application areas, such as robotics and AI-enabled human-robot interaction [32, 58, 92, 93, 97].

Considering these insights in the light of traditional CAC approaches, it seems appropriate to introduce a new paradigm. The new paradigm may be described as *Generative Context-Aware Computing* (GCAC). GCAC entails leveraging LLMs to process contextual data from sensors and other devices, thereby enhancing the acceptance and effectiveness of human-computer interactions. LLMs are well-suited for this role due to their ability to manage high-dimensional context data capturing complex and nuanced environmental and user-specific information, which are all necessary requirements for context-aware computing. Additionally, their generative capabilities allow them to infer and substitute missing data, thereby potentially maintaining robustness and adaptability in dynamic environments. This approach promises more personalized and context-aware content, improving user engagement and the overall experience with human-computer interactions such as the JITAIs in our study.

## 5.2 Limitations

This research is formative and exploratory, and clear limitations should be considered. Firstly, the generation and assessment of JITAIs were executed based on HCP-validated personas (modeling common patients) and ecologically grounded (based on prior work with stakeholder involvement) – but artificially composed – context information. The assessments of the JITAIs were based on reading and considering generated responses and not based on messages being experienced in situated daily living during live deployment.

While our study strongly indicates the viability of using LLMs like GPT-4 for creating JITAIs, further work to prove effectiveness in practice (i.e., whether the evidenced potential to motivate PA actually converts into increased PA intention formation and successful enactment, thus, crossing the intention-behavior gap [65]) is required. Our outcomes can serve to justify – also from an ethical perspective – further investment towards clinical trials in this area, e.g., in settings with CVD patients, even despite the notable degree of effort, cost, and minor inherent risk.

We made the comparison with human-generated proposed decisions and content as we assumed this is currently the best quality we can achieve, and which is common in rehabilitation settings. We did not compare performance with traditional rule-based solutions or parameter prediction based on more classical machine learning approaches, such as supervised learning from labeled datasets. As we lay out above, the ability of LLMs to process sparse and varying contextual variables and transform between semi-structured data and natural language can change the way we think about JITAIs fundamentally. This flexibility makes a clean comparison to rule-based implementations almost impossible.

Focusing solely on GPT-4, we did not evaluate other state-of-the-art LLMs. Open-source models such as the LLaMA model series [80] could be interesting candidates for future deployments, together with models that are being tailored for local execution and can fully respect crucial privacy sandboxes with federated learning (e.g., MPT-7B [50] or Phi series [34]), more recent reasoning models, including DeepSeek-R1 [9], or models trained specifically for the medical context (e.g., the Med-PaLM series [81]), as they might be more suitable concerning sensitivities and regulatory requirements in the health domain. Moreover, our results only reflect Chat-GPT's performance in the JITAI generation task at the time of data collection (14th to 21st of June 2023). Thus, outcomes may change in the future due to updates or changes in the implementation by OpenAI.

Lastly, although related work is beginning to show that fine-tuning and prompting can have a considerable impact on LLM performance [55] but also that foundation models with good prompting can often outperform domain-specialized models [42, 55, 91], we did not fine-tune and also did not experiment with more elaborate prompting strategies beyond providing factual and outcome-oriented instructions that were pre-tested with LLMs and human generators alike. Thus, further improvements could be seen from fine-tuning or applying advanced prompt engineering.

## 5.3 Implications and Future Research

Future works should compare alternative models, prompting strategies, and knowledge supplementation options more closely. Smaller, more domain-specific, and multi-modal models have the potential to more flexibly ingest and produce various input or output data formats, lead to energy savings, and, importantly, be reliably and entirely privacy-preserving, e.g., by running on user-dedicated HIPAA and GDPR-conforming secure cloud instances, or even on a user's smartphone [35]. Future studies should also focus on implementing GenAI-enabled JITAIs in practical healthcare settings, enabling further evaluation of their effectiveness in real-world environments.

In settings that potentially involve more severe consequences for triggering flawed JITAIs than fostering PA, it could also be an interesting approach to combine the LLM's ability to produce personalized and contextualized JITAI content with pre-defined rules, e.g., defined by HCPs or taken from a theory informed knowledge base via RAG. Here, it could be interesting to investigate what kind of behavior change techniques can be identified in JITAIs generated by a zero-shot model compared to an LLM that is, e.g., informed by a behavior change ontology [49].

Another intriguing aspect for future research is the exploration of the extent to which personalization and contextualization contribute to the effectiveness of JITAIs. Understanding the relative importance of personalization, e.g., as compared to viable pre-produced building blocks and content of JITAIs, could significantly influence how JITAIs are structured and delivered, potentially leading to more efficient health interventions. Moreover, the concept of incorporating a reinforcement-learning-like mechanism to enhance personalization in JITAIs presents an exciting avenue for research. Such a mechanism could involve feedback loops where patient responses to JITAIs are fed back to the LLM to continually refine and adapt the interventions based on what works for the individual. Looking beyond contributing personal preferences and feedback as part of an online RL or RL from human feedback [57] loop, further building up and considering user models offers exciting perspectives both for further improved acceptance and effectiveness of interventions. This could, e.g., be enabled by personal health knowledge graphs [25] that a GenAI model can rely on, contribute to, and expand upon. Additionally, this might even allow for scaling research into understanding the complex interplay of the high-dimensional parameter spaces that make up the composition of complex, dynamically evolving personalized interventions [76].

Further viable technical steps would include experimentation with LLMs that are fine-tuned based on the scenarios and feedback data collected from this study. For example, our qualitative findings on themes of *positive reinforcement*, *offering alternative PA options*, and *encouraging spontaneous PA for mood improvements*, which received very positive feedback over all generator groups, could inform instruction prompts for generating JITAIs.

# 6 Conclusion

In this formative study, we explored the potential of LLMs - such as GPT-4 - in creating JITAIs for cardiac rehabilitation. Our findings indicate that LLM-generated JITAIs surpassed human-generated intervention suggestions in terms of *appropriateness*, *engagement*, *perceived efficacy, and professionality*. We also demonstrated that this approach has significant adaptability across various patient personas and contexts. Furthermore, the response assessment by LayPs did not differ significantly if the LLM was identified as the producer of the JITAI, indicating an absence of critical biases regarding the potential and viability of AI systems to be the key driver in a JITAI mechanism in the sampled population. Together, these points indicate the relevant potential of GenAI with foundation models in delivering scalable, personalized, and contextualized digital health interventions [15]. Our insights highlight the potential of LLMs to transform CAC, leading us to suggest a new paradigm of *Generative Context-Aware Computing*. Despite the identified advantages, the transition from viability to efficacy and acceptability in real-world healthcare settings remains a vital area for future research. In conclusion, this study underlines the transformative potential of LLMs for digital health interventions, particularly in creating effective and personalized JITAIs for fostering behavioral change, clearly warranting investment into further study of the effectiveness of situated and longer-term deployments.

## ACKNOWLEDGMENTS

We thank all our study participants. We also thank Dr. Daniela Wurhofer and Dr. Stefan Tino Kulnik, as well as the wider teams behind the iterative development process of the aktivplan application in which this exploratory work is grounded for motivation and which produced the original personas that were adapted for this work. We also thank various participants of the GenAI for Digital Health Interventions workshop held in Salzburg for the early discussion around the research aim of this work.

### Author Contributions

DH acted as the lead author of the study and implemented and executed most of the study. JS acted as the senior author, providing mentorship and guidance throughout the project. JS and DK developed the original research concept and collaborated on iterating it with DH throughout the study duration. DH, DK, SG, DHo, MS, GT, JN, CB, AS, and JS all contributed to drafting and improving the writing of this article. MS and GT contributed to checking the fidelity of personas and scenarios and – together with JN supported writing the medical / health-relevant sections of the draft. AS particularly supported contextualization in context-aware computing and implications and relevance for HCI.